\documentclass[preprint,prl,showpacs,floatfix]{revtex4}

\usepackage{graphicx}

\begin{document}

\title
{
Discreteness-Induced Slow Relaxation in Reversible Catalytic Reaction Networks
}

\author
{
Akinori Awazu$^1$ and Kunihiko Kaneko$^{2,3}$
}

\affiliation
{
$^1$Department of Mathematical and Life Sciences, Hiroshima University, Kagami-yama 1-3-1, Higashi-Hiroshima 739-8526, Japan.
}

\affiliation
{
$^2$Department of Basic Science, University of Tokyo, Komaba, Meguro-ku, Tokyo 153-8902, Japan.
}

\affiliation
{
$^3$ERATO Complex Systems Biology, JST, Komaba, Meguro-ku, Tokyo 153-8902, Japan.
}

\date{\today}

\begin{abstract}
Slowing down of the relaxation of the fluctuations around
equilibrium is investigated both by stochastic simulations and by
analysis of Master equation of reversible reaction networks
consisting of resources and the corresponding products that work
as catalysts. As the number of molecules $N$ is decreased, the
relaxation time to equilibrium is prolonged due to the deficiency
of catalysts, as demonstrated by the amplification compared to
that by the continuum limit. This amplification ratio of the
relaxation time is represented by a scaling function as $h = N
\exp(-\beta V)$, and it becomes prominent as $N$ becomes less than
a critical value $h \sim 1$, where $\beta$ is the inverse
temperature and $V$ is the energy gap between a product and a
resource.
\end{abstract}


\maketitle

\section{I. Introduction}

The study of reaction processes in catalytic reaction networks is
generally important to understand the dynamics and fluctuations in
biochemical systems and their functionality. Obviously,
understanding the generic features of equilibrium characteristics
and relaxation to equilibrium is the first step toward gaining
such an understanding. Indeed, such reaction systems often exhibit
anomalous slow relaxation to equilibrium due to some kinetic
constraints such as diffusion-influenced (limited)
reaction\cite{DLR} and formations of transient Turing
patterns\cite{AK2}. In this paper, we consider a novel mechanism
to realize such slow relaxation in catalytic reaction networks,
where the discreteness in molecule number that may reach zero
induces drastic slowing down.

Most intra-cellular reactions progress with the aid of catalysts
(proteins), whereas catalysts have to be synthesized as a
result of such catalytic reactions. Indeed, reaction dynamics in
catalytic networks have been extensively investigated. In most
such studies, a limiting case with a strong non-equilibrium
condition was assumed by adopting a unidirectional reaction
process (i.e., by neglecting backward reactions). To understand
the basic properties of biochemical reactions, however, it is
important to study both equilibrium and non-equilibrium
characteristics by including forward and backward reactions that
satisfy the detailed balance condition. Such a study is not only
important for statistical thermodynamics but it also provides some
insight on the regulation of synthesis or degradation reactions
for homeostasis in cells.

Recently, we discovered a slow relaxation process to equilibrium,
which generally appears in such catalytic reaction networks, and
proposed
 "chemical-net glass" as a novel class of nonequilibrium phenomena.
In this case, relaxation in the vicinity of equilibrium is
exponential, whereas far from it, much slower logarithmic
relaxation with some bottlenecks appears due to kinetic
constraints in catalytic relationships\cite{AK3}. In this study,
we adopted continuous rate equations and assumed that the molecule
number is sufficiently large.

In biochemical reaction processes, however, some chemical species
can play an important role at extremely low concentrations of even
only a few molecules per cell\cite{cell2,cell3,cell4}. In such
systems, fluctuations and discreteness in the molecule number are
important. Indeed, recent studies by using a stochastic simulation
of catalytic reaction networks have demonstrated that the
smallness in the molecule number induces a drastic change with
regard to statistical and spatiotemporal behaviors of molecule
abundances from those obtained by the rate equation, i.e., at the
limit of large molecule
numbers\cite{togashi1,ookubo,AK1,AK11,mif1,mif2,Solomon,togashi3,marion,zhdanov,Dau,Kaneko-Adv,Furusawa,Furusawa2}.
In these studies, the strong nonequilibrium condition is assumed
by taking a unidirectional reaction.

Now, it is important to study the relaxation process to
equilibrium by considering the smallness in the molecule number.
Does the discreteness in molecule number influence the equilibrium
and relaxation behaviors?  Is the relaxation process slowed down
by the smallness in the molecule number? To address this question,
we have carried out several simulations of the relaxation dynamics
of random catalytic reaction networks by using stochastic
simulations. Numerical results from several networks\cite{SAK1,SAK2}
suggest that the relaxation time is prolonged drastically when the
number of molecules is smaller. The increase from the continuum
limit is expressed by the factor $\exp(\beta \delta E)$, where
$\delta E$ is the additional energy required to pass through the
bottleneck due to the discreteness in molecule number and $\beta$
is the inverse temperature.

In this paper, we analyze such slowing down of a reaction process
to equilibrium that is induced by the smallness in molecule
numbers. Instead of taking complex reaction networks, we choose
simple networks or network motives to estimate the relaxation time
analytically. In fact, complex networks are often constructed by
combining a variety of simple network motives with simple branch
or loop structures. We focus on the relaxation dynamics of
reversible catalytic reaction systems with such simple network
motives as a first step toward understanding the general
relaxation properties in complex catalytic reaction networks.

In section II, we introduce two network motives, where the
synthesis of a product from resource molecules (and its reverse
reaction) is catalyzed by one of the other products. Here, we note
that some specific network motives may exhibit incomplete
equilibration when the molecule number decreases, and the average
chemical concentration in the steady state deviates from the
equilibrium concentration derived by the continuous rate
equations.

In section III, we show relaxation characteristics from the
stochastic simulations. The relaxation of the fluctuation around
the steady state slows down as the molecule number is decreased
below a critical value.  This increase is represented by a scaling
function by using $h = N \exp(-\beta V)$, where $N$ is the
molecule number and $V$, the energy gap between a product and a
resource. In section IV, we present an analytic estimate for this
relaxation suppression due to the smallness in molecule number by
using a suitable approximation for Master equation. In section
V, we present a summary and discuss the generality of our results.

\section{II. Models}

\begin{figure}
\begin{center}
\includegraphics[width=6.0cm]{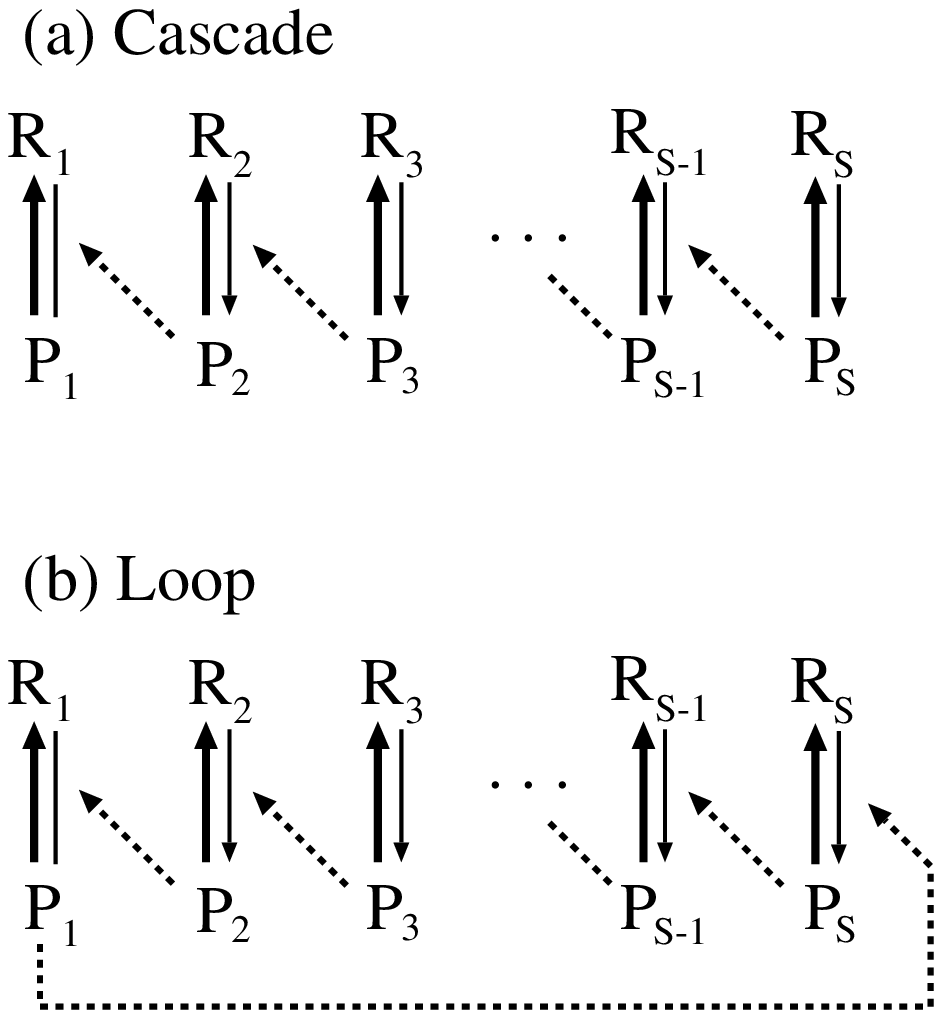}
\end{center}
\caption{Illustration of (a) cascade system and (b) loop system.
Solid arrows indicate reaction paths (their width indicates the
transition tendency) and dashed arrows indicate catalyzation.}
\end{figure}

Here, we consider reversible catalytic reaction systems with two
simple network structures, cascade system and loop system, as
shown in Fig. 1, which may function as network motives for complex
reaction networks. These systems consist of $2S$ chemical species,
which are Product $P_i$ and Resource $R_i$ with $i = 1, 2, ...,
S$. Here, each product chemical can catalyze at most one of the
other Resource-Product reactions, whereas each reaction is
catalyzed at most by some product. (Instead, we can interpret that
there exist $S$ chemical species with excited and non-excited
states, and chemicals in an excited state can catalyze an
excitation reaction of one of the other molecules.)

If all chemicals are catalyzed by one of them, we can renumber
$P_i$ and $R_i$ for $i = 1, 2, ..., S-1$ and write the reaction
 as

$P_i + P_{i+1} \rightleftharpoons^{k_{P_i,R_i}}_{k_{R_i,P_i}} R_i + P_{i+1}$,
\\[1ex]
where $P_S + P_{1} \rightleftharpoons^{k_{P_S,R_S}}_{k_{R_S,P_S}}
R_S + P_{1}$, which leads to the loop system (b).  When there
exist a reaction that is not catalyzed, the cascade system in
Fig.1a) is obtained where
 $P_S \rightleftharpoons^{k_{P_S,R_S}}_{k_{R_S,P_S}} R_S$.
(Neglecting cases in which some pair of resource and product is
totally disconnected from others, the loop and cascade systems are
the only possibilities).

The rates of forward ($k_{P_i,R_i}$) and backward ($k_{R_i,P_i}$)
reactions are set so that they satisfy the detailed balance
condition. We assume that the energy of the chemical $P_i$ is
larger than that of $R_i$, and we set $k_{P_i,R_i} = 1$ and
$k_{R_i,P_i} = \exp(-\beta V_i)$, where $V_{i}$ is the energy gap
between $P_i$ and $R_i$ and $\beta$ is the inverse temperature. We
define $p_i$ and $r_i$ as the number of molecules of the chemical
species $P_i$ and $R_i$, respectively. We fix the total number of
molecules as $SN$, and $p_i + r_i = N$ holds for each $i$. The
state of the system is represented by a set of numbers $(p_1, p_2,
... , p_S)$.

In both the systems, it is noted that for $N \to \infty$ (i.e.,
the continuous limit), $<p_i> \to p_i^{eq} = \frac{N e^{-\beta
V_i}}{1+e^{-\beta V_i}}$ and $<r_i> \to r_i^{eq} =
\frac{N}{1+e^{-\beta V_i}}$ holds at the equilibrium distribution,
which is reached at $t \to \infty$.

For finite $N$, however, there is a difference between the
distribution of the cascade and the loop systems. In the cascade
system, the average of the equilibrium chemical concentrations are
identical to the continuum limit, and are given by
 $<p_{i}> = \frac{Ne^{-\beta V}}{1+e^{-\beta V}}$, that is, they
 are independent of $N$ and $\beta$.   This is because all the states
$(p_1, p_2, ... , p_S)$ ($0 \leq p_i\leq N$) are connected by
reactions and the above equilibrium distribution is only the
stationary solution for Master equation.

On the other hand, in the loop system, there is a deviation in the
steady chemical concentration from the continuum limit, which
becomes more prominent as $N$ becomes smaller. This is because the
state $(p_1, p_2, ..., p_S) = (0, 0, ..., 0)$ cannot be reached
from other states, whereas the state cannot move to any other
states. Hence, the steady distribution from the initial conditions
without $(p_1, p_2, ..., p_S) = (0, 0, ..., 0)$ deviates from the
continuum limit. This deviation becomes prominent as $N$ becomes
smaller. For example, for $N=1$ and $V_i=V$, the distribution from
the initial condition without $(p_1, p_2, ..., p_S) = (0, 0, ...,
0)$ is given by $<p_{i}> = \frac{e^{-\beta V}(1+e^{-\beta
V})^{S-1}}{(1+e^{-\beta V})^S - 1}$. Note that $<p_{i}>$ tends to
$1/S$ with an increase in $\beta$.

\section{III. Simulation results}

In this section, we present the results of stochastic simulations
and show the dependence of the relaxation process on the number of
molecules $N$ and the inverse temperature $\beta$. For simplicity,
we consider $V_i$ to be uniform for all species ($= V$); however,
this assumption can be relaxed.

Numerical simulations are carried out by iterating the following
stochastic processes. (i) We randomly pick up a pair of molecules,
say, molecule 1 and 2. (ii) Molecule 1 is transformed with its
reaction rate (if it is P, it is transformed to R, and vice versa)
if molecule 2 can catalyze the reaction of molecule 1. In the
cascade case, there is a reaction that progresses without a
catalyst, and in this case, if molecule 1 is the one that reacts
without a catalyst, then it is transformed with the reaction rate
independently of 2. Here, a unit time is defined as the time span
in which the above processes for catalytic reactions are repeated
$SN$ times. In each unit time, each molecule is picked up on
average to check if the transformation occurs.

In the following, we focus on the behavior of the system after a
sufficiently long time from the initial time where the numbers of
each molecule $p_i$ and $r_i$ are set randomly from $[0,N]$ under
the constraint $p_i + r_i = N$ and $(p_1, p_2, ..., p_S) \neq  (0,
0, ..., 0)$.

\begin{figure}
\begin{center}
\includegraphics[width=7.0cm]{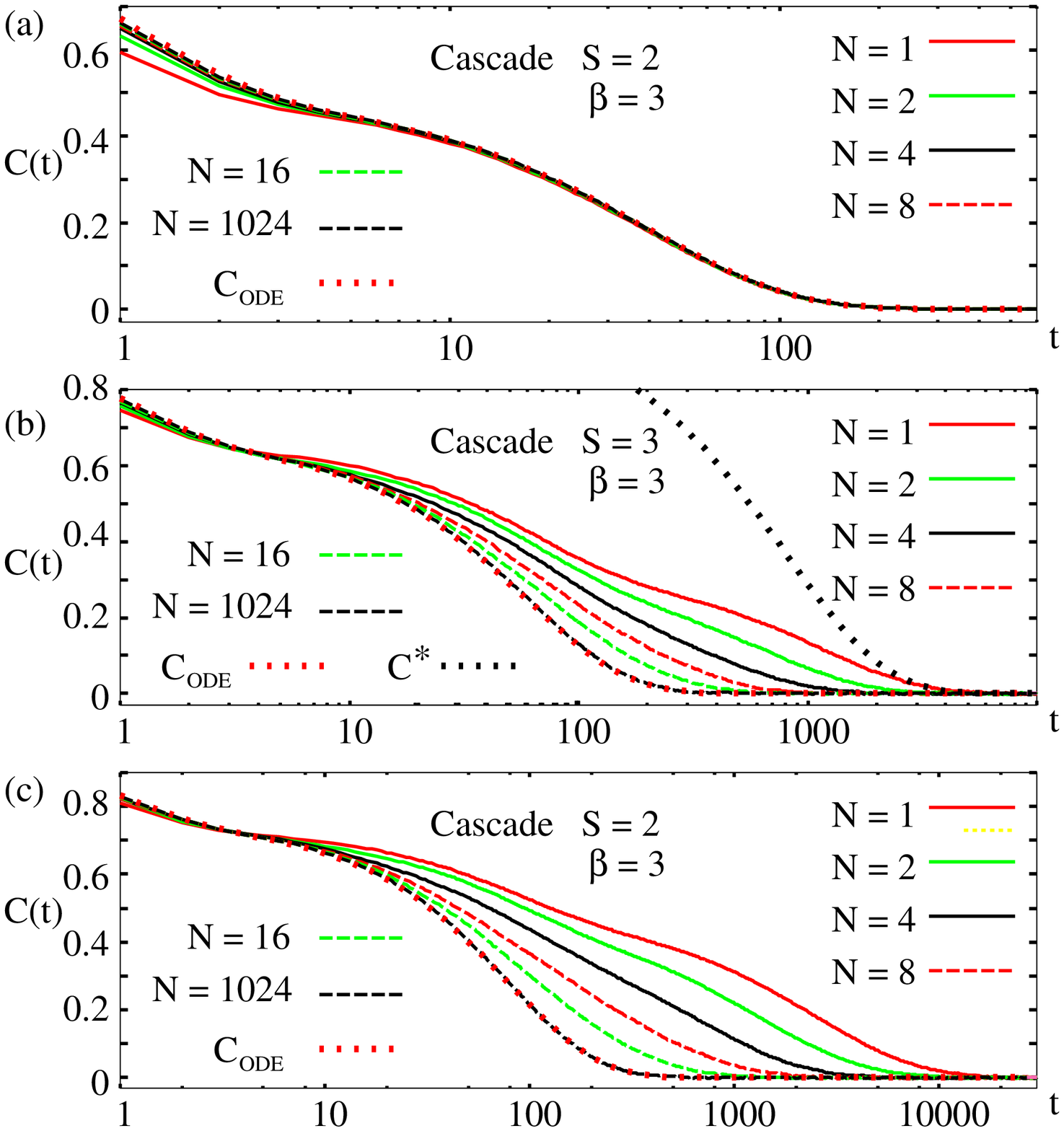}
\includegraphics[width=7.0cm]{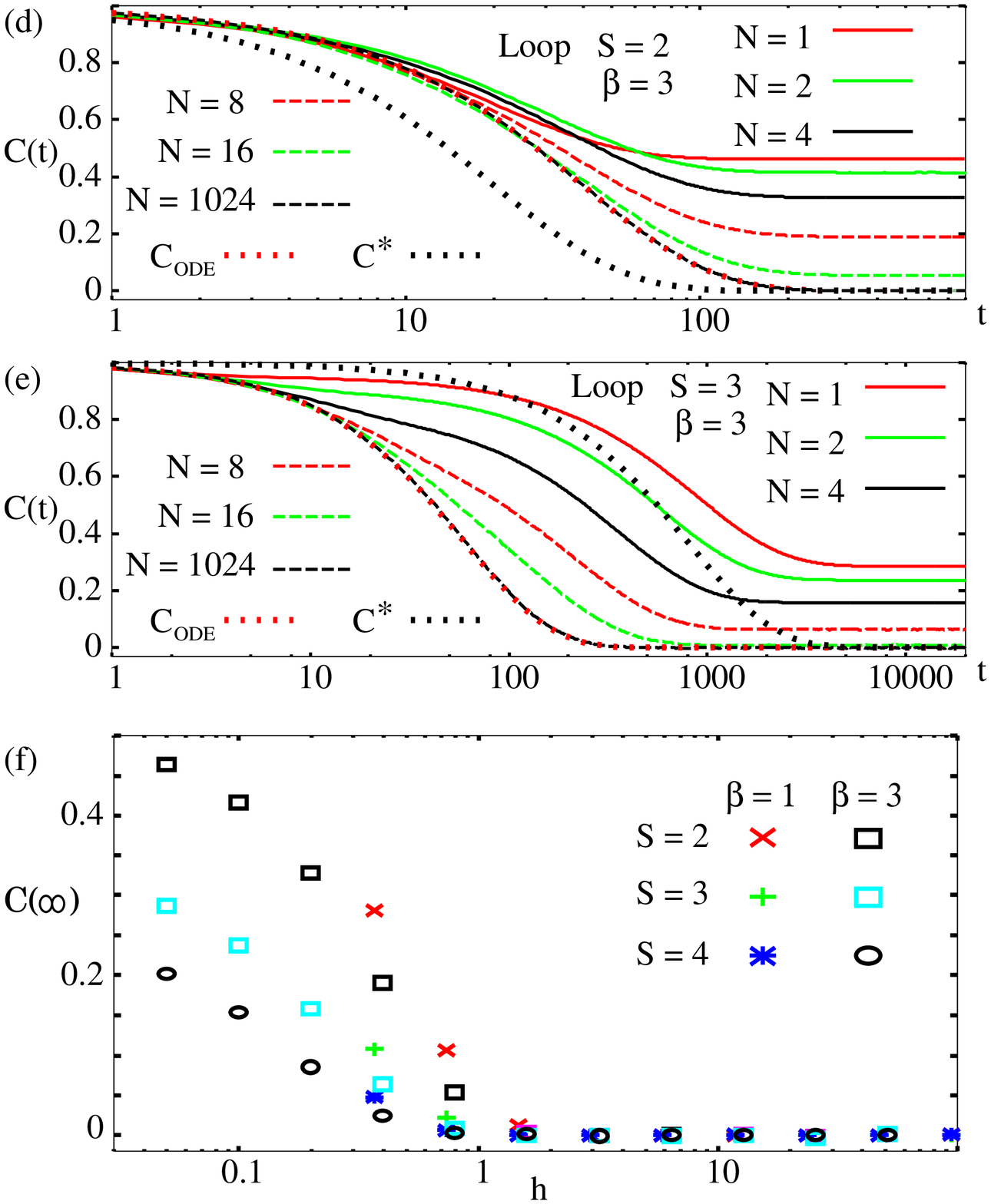}
\end{center}
\caption{$C(t)$ of cascade systems with (a) $S = 2$,  (b) $S = 3$,
and (c) $S=4$, and loop systems with (d) $S = 2$ and (e) $S = 3$
for several N with $\beta = 3$. (f) $C(\infty)$ as a function of
$h$ in loop systems for several $\beta$ and $S$. $C_{ODE}$
indicates the auto-correlations given by Eq. (4) in (a)-(c) and
Eq. (3) in (d) and (e). $C^{*} = \exp(-e^{-\beta V}t)$ in (d), and
$C^{*} = \exp(-\frac{e^{-2\beta V}}{2}t)$ in (b) and (e) with
$\beta = 3$ ($V = 1$).}
\end{figure}

Figure 2(a)-2(e) show the auto-correlation functions of the
deviation from the equilibrium concentration of the cascade system
((a)-(c)) and the loop system ((d) and (e)) for some $S$ and N
with $\beta = 3$, defined by $C(t) = c(t)/c(0)$ and $c(t) =
<\sum_i [(p_i(t) - p_i^{eq})(p_i(0) - p_i^{eq}) +
(r_i(t)-r_i^{eq})(r_i(0) - r_i^{eq})]>$. As already discussed,
$C(\infty) \to 0$ in the cascade system whereas $C(\infty) > 0$
for small $N$. The value $C(\infty)$ starts to deviate when $h=N
e^{-\beta V}$ becomes less than 1. Hence, we have plotted
$C(\infty)$ of the loop system as a function of $h$ in Fig.2(f)
for $\beta = 1$ and $3$. As shown, $C(\infty) > 0$ holds for $h <
1$ independently of $\beta$. On the other hand, in both systems,
the relaxation to the final state with $C(t) = const.$ for small
$N$ is drastically slowed down as compared to that for large $N$
when $S > 2$, whereas the relaxation for small $N$ is faster when
$S = 2$.

\begin{figure}
\begin{center}
\includegraphics[width=8.0cm]{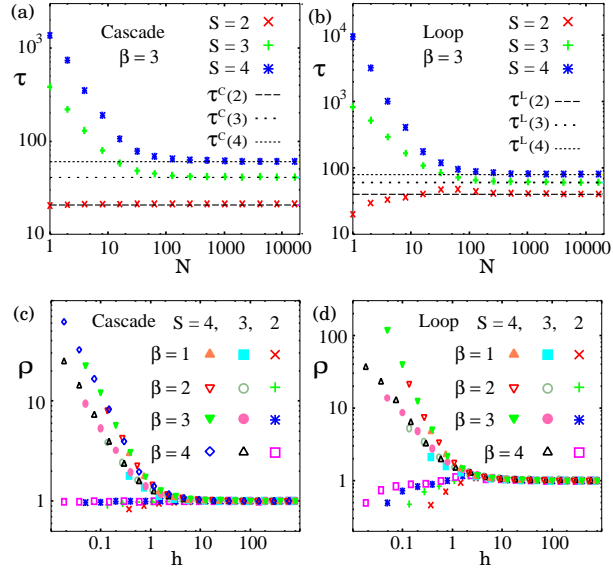}
\end{center}
\caption{$\tau$ as a function of $N$ in (a) cascade system and (b)
loop system, and $\tau^{C}(S)$ and $\tau^{L}(S)$
for $\beta=3$ and $S= 2, 3, 4$. $\rho$ as a function of $h$ in (c)
cascade system (d) loop system with $S= 2, 3, 4$ for several
$\beta $}
\end{figure}

To observe the dependence of the relaxation time on $N$, we
measured the integrated relaxation time defined as $\tau =
\int_0^{\infty} \frac{C(t)-C(\infty)}{1-C(\infty)}dt$. Figure 3(a)
and (b) show $\tau$ as a function of $N$ for $\beta=3$ with $S= 2,
3, 4$ for the (a) cascade system and (b) loop system. For $S \geq
3$, the relaxation time $\tau$ increases by several orders of
magnitude with a decrease in $N$ in both systems. On the other
hand, $\tau$ for $S = 2$ does not exhibit any drastic change with
the decrease in $N$ in both systems.

This prolongation of $\tau$ for $S>2$ becomes more prominent as
$\beta$ is increased. From several data, $\tau$ is suggested to
increase as a function of $\exp(\beta V$). Combining $N$ and
$\beta$ dependencies, we introduce a parameter $h=N \exp(-\beta
V)$. The discreteness effect is dominant when $h=N \exp(-\beta V)$
is less than unity. Figure 3(c) and 3(d) show $\rho = \tau /
\tau_{N \to \infty}$ as a function of $h$ for the (c) cascade
system and (d) loop system for several values of $\beta$ and $S=
2, 3, 4$. For $S>2$, the deviation of $\rho$ from the continuum
limit ($\rho=1$) becomes prominent when $h$ is below unity in both
systems. The increase in $\rho$ appears to become steeper with an
increase in $S$. On the other hand, $\rho$ for $S = 2$ does not
exhibit a drastic increase with a decrease in $h$.

\section{IV. Origin of slow relaxations and crossover}

\subsection{A. Relaxation processes for $N \to \infty$ and $N=1$}

Now, we analytically estimate the enhancement in relaxation time
and explain its representation in the form $h = N\exp(-\beta V)$.
For this purpose, we compare the estimate by Master equation
analysis for small $N$ and compare it with that from the continuum
limit $N \to \infty$.

In the continuum limit, the reaction dynamics are represented by
the following rate equation:
\begin{equation}
\dot{x_{i}} = x_{c}[ e^{-\beta V} (\frac{1}{S} - x_{i}) - x_{i}]
\end{equation}
with $x_{i} = p_i / SN$. Here, $x_c = 1$ for $i=S$ in the cascade
system and $x_c = x_1$ for $i=S$ in the loop system. In both
systems, $x_{i} \to x_i^{eq} = \frac{e^{-\beta V}}{S(1+e^{-\beta
V})}$ holds for $t \to \infty$. When the deviation from
equilibrium $\delta x_i=x_{i} - x_i^{eq}$ is small, its evolution
for the loop systems obeys the following linearized equation
\begin{equation}
\dot{\delta x_{i}} = -\frac{e^{-\beta V}}{S} \delta x_{i}.
\end{equation}
For the cascade system, this equation is also valid for the
elements $i \neq S$, whereas $\dot{\delta x_{S}} = -\delta x_{S}$.
Then, the auto-correlation function of a small fluctuation of
$p_i$ around $p_i^{eq}$ is obtained as
\begin{equation}
C(t) = \exp(-\frac{e^{-\beta V}}{S}t)
\end{equation}
for the loop system, and
\begin{equation}
C(t) = \frac{1}{S}\exp(-t) + \frac{S-1}{S}\exp(-\frac{e^{-\beta V}}{S}t)
\end{equation}
for the cascade system. Indeed, these agree quite well with the
simulation results for a sufficiently large $N$ (e.g., $N= 1024$
in Fig. 2.). Thus, the characteristic time of the relaxation is
estimated as $\tau^{L}(S) \sim Se^{\beta V}$ for the loop system
and $\tau^{C}(S) \sim \frac{1}{S}+(S-1)e^{\beta V}$ for the
cascade system, which are consistent with the simulation results
shown in Fig. 3.

As the other extreme limit, consider the case with $N=1$. In this
case, the relaxation dynamics are dominated by a completely
different process induced by the absence of catalysts whose number
can often go to zero. In such cases, states are trapped at some
local energy minimum that appears due to the deficiency of
catalysts. Then, the hopping processes among them play an
important role in the relaxation dynamics, as shown below. In the
following, we focus the cases with $S = 2$ and $S = 3$ to clarify
that such an effect is induced by discreteness in the molecule
number. Note that, as shown in the last section, the behavior for
$S \ge 3$ is distinct from that for $S=2$; in the former case, the
relaxation time is enhanced by the decrease in $N$, in contrast to
the latter case.

\begin{figure}
\begin{center}
\includegraphics[width=7.0cm]{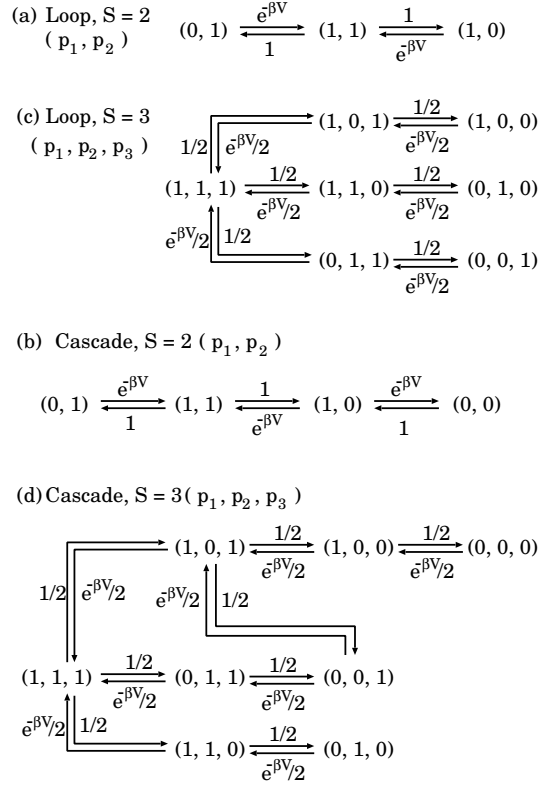}
\end{center}
\caption{(a) Illustration of transition diagrams of (a) loop
system with $S=2$, (b) loop system with $S=s$, (c) cascade system
with $S=2$, and (d) cascade system with $S=3$, where arrows
indicate possible transitions and the values next to them specify
the transition ratios.}
\end{figure}

First, we study the loop system. When $S = 2$, the system realizes
3 states from the initial conditions---$(p_1, p_2) = (1,0)$,
$(0,1)$, and $(1,1)$---as shown in Fig. 4(a).
Then, we estimate the time of the transition between $(1,0)$ and $(0,1)$.
First, the transition rate from the state $(1,0)$ to $(1,1)$ is estimated
as follows: for this  transition, a pair of molecules from the product of
the first species and the resource of the second species has to be chosen. 
This probability is given by $\frac{1}{2}\frac{1}{2-1}$, while
the reaction rate is given by $e^{-\beta V}$.  Hence the rate is given by
$2 \cdot \frac{1}{2}\frac{1}{2-1}e^{-\beta V}= e^{-\beta V}$. Thus, the
characteristic time of the correlation of each $p_i$ is given by
$\sim e^{\beta V}$, which is consistent with the results shown in
Fig. 2(d).
%

On the other hand, for $S = 3$, the system realizes 7
states---$(p_1, p_2, p_3) = (1,0,0)$, $(0,1,0)$, $(0,0,1)$
,$(1,1,0)$, $(1,0,1)$, $(0,1,1)$, and $(1,1,1)$---as shown in Fig.
4(b). The characteristic time of the correlation of each $p_i$ is
given by the transition time among the three branches including
lowest-energy states, $(1,0,1)$ - $(1,0,0)$,
$(1,1,0)$ - $(0,1,0)$, and  $(0,1,1)$ - $(0,0,1)$.
Here, in order to hop from one branch to another, the system must go
through the highest-energy state $(1,1,1)$, due to the
restriction by the catalytic relation. Now, we define the probability that
the states in the branch $(1,0,1)$ - $(1,0,0)$ are realized as $Q_{1,0,0}$.
Then, the probability to realize the state $(1,0,1)$ is given by
$\frac{e^{-\beta V}}{1+e^{-\beta V}}Q_{1,0,0}$.
Here, the transition rate from $(1,0,1)$ to $(1,1,1)$ is given by
$\frac{e^{-\beta V}}{2}$. Then, the probability current from the
the branch $(1,0,1)$ - $(1,0,0)$ is estimated by
$\sim \frac{e^{-\beta V}}{2}\frac{e^{-\beta V}}{1+e^{-\beta V}}Q_{1,0,0} \sim \frac{1}{2}e^{-2\beta V}Q_{1,0,0}$ ($e^{-\beta V} << 1$). Because of the symmetry among
the catalytic reactions, the
probability currents from the other branches are obtained in the same
way, to get the same form. Thus, the escape rate from each branch is
estimated by $\sim \frac{1}{2}e^{-2\beta V}$, and the characteristic time
of the correlation of each $p_i$ is estimated as $\sim 2e^{2\beta V}$.
Because the relaxation time in the continuum limit is proportional to
$\exp(\beta V)$, the deviation $\rho$ from it increases with $\exp(\beta V)$,
which is consistent with the results shown in Fig. 2(e). Thus, the
enhancement of the relaxation time from the continuous case is explained.

Essentially the same argument is also valid for the cascade
systems. When $S = 2$, the system can realize transitions among 4
states---$(0,1) - (1,1) - (1,0) - (0,0)$---as shown in Fig. 4(c).
Here, $(0,1)$ is a metastable state and $(0,0)$ is the
lowest-energy state. The relaxation is characterized by the escape
rate from a metastable state, which is given by $\sim e^{-\beta
V}$. Thus, the characteristic time of the correlation of each
$p_i$ is given by $\sim e^{\beta V}$.

On the other hand, for $S = 3$, the system realizes 8
states--$(p_1, p_2, p_3) = (0,0,0)$, $(1,0,0)$, $(0,1,0)$,
$(0,0,1)$ ,$(1,1,0)$, $(1,0,1)$, $(0,1,1)$, and $(1,1,1)$---as
shown in Fig. 4(d). The slowest characteristic time of the relaxation
is given by the transition time from the branch, $(1,1,0)$ - $(0,1,0)$
since the system must go through the highest-energy state $(1,1,1)$,
which is a limiting process for this case. Then,
in a manner similar to the loop system with $S=3$, the
characteristic time is obtained as $\sim 2e^{2\beta V}$. This
gives the characteristic time of the slowest motions of the
system. This estimation fits well with the numerical result shown
in Fig. 2(b).

\subsection{B. $N$, $\beta$ dependencies of $C(\infty)$ and relaxation time}

Next, we extend the argument of the last subsection to analyze the
$N$ and $\beta$ dependencies of $C(\infty)$ and the relaxation
time in greater detail. In particular, we explain why $h =
N\exp(\beta V) \sim 1$ gives a critical value and how the
amplification of relaxation time depends on $h$ for $h<1$. Because
of the simplicity due to the symmetry in the catalytic
relationship, we only study loop systems; however, the argument
presented below can be extended to cascade systems.

Figure 5(a) shows the transition diagram of the loop system with
$S = 2$, where each circle indicates each state $(p_1,p_2)$ and
the arrows indicate possible transitions. Generally, for any
values of $S$, the transition rate from a state $(p_1,p_2,...,
p_i=n, p_{i+1}, ...,p_S)$ to a state $(p_1,p_2,..., p_i=n+1,
p_{i+1}, ...,p_S)$ per unit time is estimated as follows. For this
transition, a pair of molecules from the resource of the $i$th
species ($R_i$) and the product of the $(i+1)$th species
($P_{i+1}$) has to be chosen. This probability is given by
$\frac{N-p_i}{SN}\frac{p_{i+1}}{SN-1}$, and the reaction rate is
given by $e^{\beta V}$. Hence, the transition rate per unit time
is given by $W^i_{n \to n+1} = \frac{(N-n)p_{i+1}}{SN-1}e^{-\beta
V}$. Similarly, the transition rate in the opposite direction is
given as $W^i_{n+1 \to n} = \frac{(n+1)p_{i+1}}{SN-1}$. If the
molecule number is so large or $\beta$ is so small that $h =
Ne^{-\beta V} >> 1$, $W^i_{n \to n+1} > W^i_{n+1 \to n}$ holds for
small $n$ and $W^i_{n \to n+1} < W^i_{n+1 \to n}$ holds for large
$n$. Then, the dominant states of the system are located in an
intermediate region in the phase space $[0,N]$. For example, the
blue region in Fig. 5(a) indicates such dominant states for $S=2$.

Now, we define the probability that $p_i=n$ as $Q_n^i$, and the joint
probability to realize $p_i=n$ and $p_{i+1}=m$ as $Q_{n,m}^i$. Here,
$Q_n^i = \sum_{m=0}^NQ_{n,m}^i$ and $Q_{n,m}^i = Q_n^iQ_m^{i+1}$. Then the time evolution of $Q_{n,m}^i$ follows
\begin{equation}
\dot{Q}_{n,m}^i = \frac{m}{SN-1}[ (N-(n-1))e^{-\beta V}Q_{n-1,m}^i+ (n+1)Q_{n+1,m}^i -nQ_{n,m}^i-(N-n)e^{-\beta V}Q_{n,m}^i ].
\end{equation}
Then, we obtain
\begin{equation}
\dot{Q_n^i} = \frac{<p_{i+1}>}{SN-1}[ (N-(n-1))e^{-\beta V}Q_{n-1}^i+ (n+1)Q_{n+1}^i -nQ_n^i-(N-n)e^{-\beta V}Q_n^i ],
\end{equation}
where $<p_i> = \sum_{n=0}^NnQ_n^i$ ($<p_{i+1}> =
\sum_{m=0}^NmQ_m^{i+1}$). Using this equation, we obtain the time
evolution of $<p_i>$ as
\begin{equation}
\dot{<p_i>} = \frac{<p_{i+1}>}{SN-1}[-<p_i>+(N-<p_i>)e^{-\beta V}]
\end{equation}
This implies that $x_{i} = <p_i> / SN$ obeys equation (1) for a
sufficiently large value of $N$.

On the other hand, if $N$ is so small or $\beta$ is so large that
$h << 1$, $W^i_{n \to n+1} << W^i_{n+1 \to n}$ holds for all $i$
and $n$. Thus, $p_i$ for all $i$ tend to decrease to $0$. Then,
there exist $S^N$ metastable states---$(n,0,0,...,0)$,
$(0,n,0,...,0)$, ... , $(0,0,...,0,n,0,..,0)$, ... , and
$(0,0,0,...,n)$ ($1 \leq n \leq  N$). Among them, the following
$S$ states, $(1,0,0,...0)$, $(0,1,0,...0)$, ..., and
$(0,0,0,...1)$, have the lowest energy. For example, in the cases
with $S=2$, the states $(0, p)$ and $(q, 0)$ ($p, q \ne 0$) are
metastable states and $(1,0)$ and $(0,1)$ are the lowest-energy
states.

It should be noted that the lowest-energy states are the dominant
states for $h << 1$. The probability to realize these
lowest-energy states tends to $1/S$ with an increase in $\beta$.
Thus, with the increase in $\beta$, $<p_i>$ approaches $1/S$ for
small $N$, which indicates $C(\infty) =const. > 0$ for small $N$
and large $\beta$.

Moreover, for $h << 1$, the transitions among lowest-energy states
contribute dominantly to the relaxation process. Then, we estimate
the characteristic time of the fluctuations of the system for $h
<< 1$ by considering the transition processes from one lowest-energy
states such as $(0, 0,...,0, p_j =1, 0 ...,0)$ to the other lowest-energy
states such as $(0, 0...,0, p_j = 0, 0, p_{j'} = 1, 0, ...,0)$.
In the following, we consider only the cases with
$S=2$ and $S=3$.  We only focus on the dynamics of $p_j$
under the constraint that $p_{j}$ has only $0$ or $1$, because $h<<1$.

\begin{figure}
\begin{center}
\includegraphics[width=7.0cm]{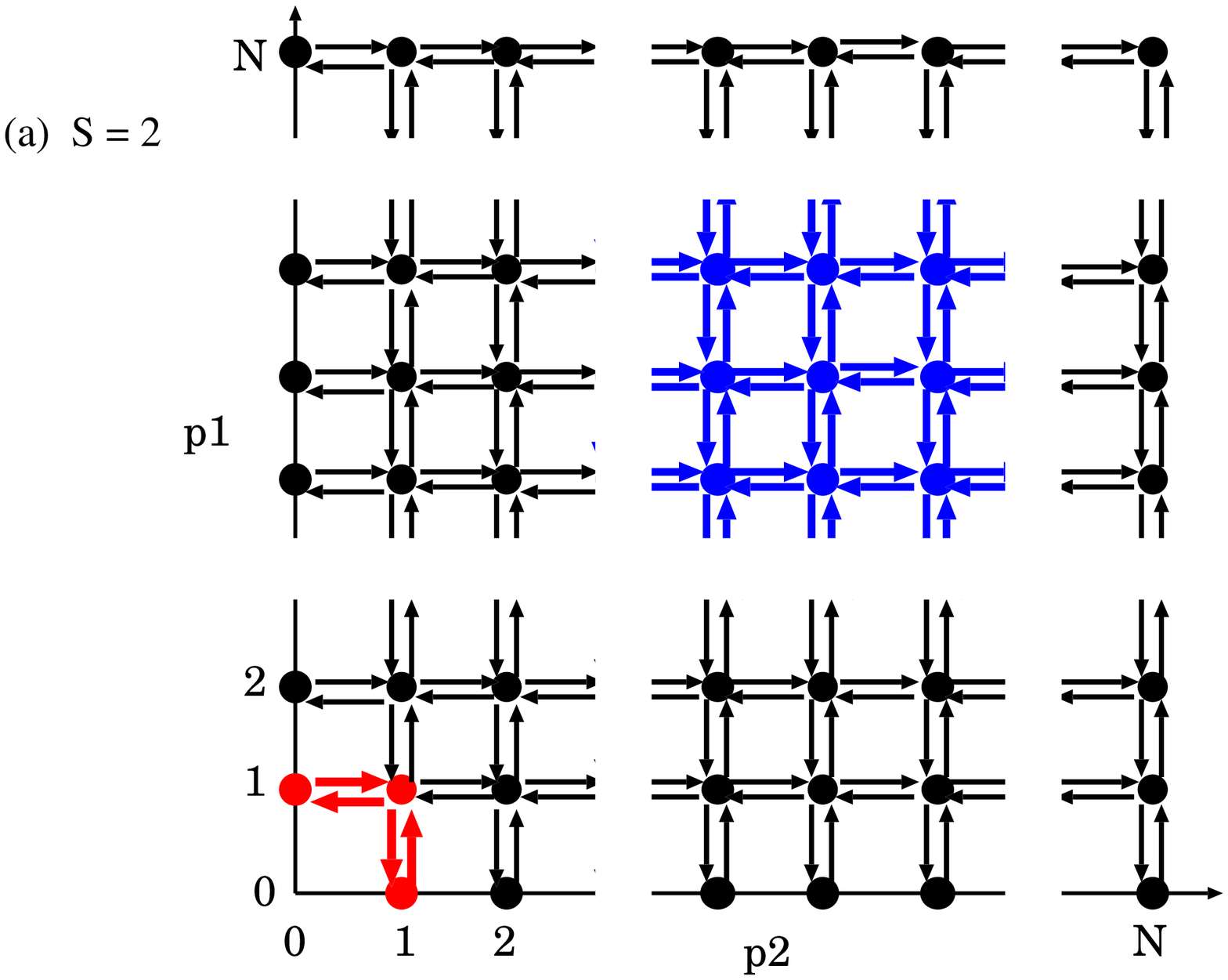}
\\[2ex]
\includegraphics[width=8.0cm]{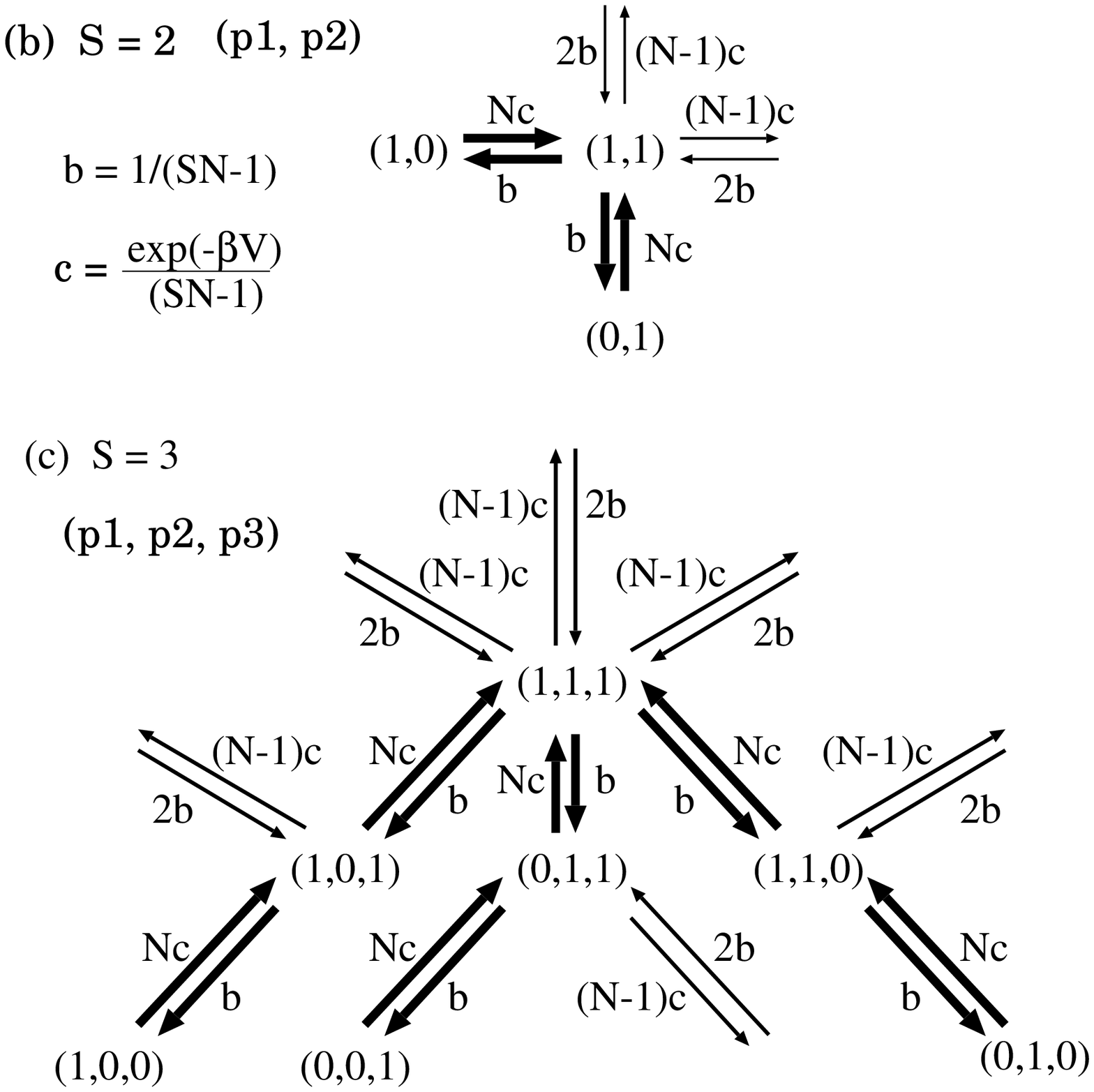}
\end{center}
\caption{
(a) Illustration of the transition diagrams for $S=2$, and
effective transition diagrams for (b) $S=2$ and (c) $S=3$, where
bold arrows indicate the focused transitions in the text.
}
\end{figure}

First, consider the case with $S=2$. Figure 5(b) shows a detailed
transition diagram around the region where $p_i$ ($i = 1, 2$) are
only $0$ or $1$. The escape rate from $(1,0)$ and $(0,1)$ are given by
$\sim \frac{N}{2N-1}e^{-\beta V}$. Thus, the
characteristic time of the correlation of each $p_i$ is given by
\begin{equation}
\tau^{L}_d(2) \sim \frac{2N-1}{N} e^{\beta V},
\end{equation}
which is consistent with the results shown in Fig. 6(a).

\begin{figure}
\begin{center}
\includegraphics[width=8.0cm]{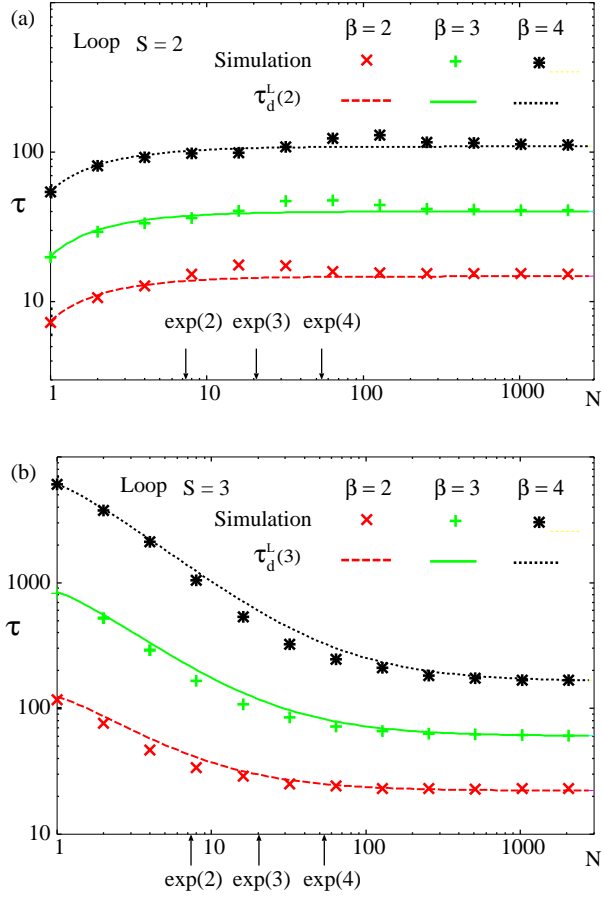}
\end{center}
\caption{Relaxation time $\tau$ obtained from simulations (points)
and its approximate analytical expression $\tau^{L}_d(S)$ (curves)
estimated in the text. Plotted as a function of $N$ for the loop
systems with (a) $S = 2$ and (b) $S = 3$ with $\beta= 2, 3, 4$.
The analytical expression agrees with the simulation data both for small $N$,
and for large
$N$, where $\tau$ approaches a constant value expected from the rate
equation.  The crossover occurs at around
$h=Ne^{-\beta V}\sim1$ ($N \sim \exp(2)$, $\exp(3)$, and $\exp(4)$ for
$\beta = 2$, $3$, and $4$.).}
\end{figure}

Next, we study the case with $S=3$. The transition diagram of the
states $(p_1,p_2,p_3)$ is shown in Fig.5(c) when $p_i$ ($i = 1, 2, 3$)
take only $0$ or $1$. Similar to the $N=1$ case, the characteristic time
of the transition among the three branches including lowest-energy states,
$(1,0,1)$ - $(1,0,0)$, $(1,1,0)$ - $(0,1,0)$, and  $(0,1,1)$ - $(0,0,1)$
through the state $(1,1,1)$ is considered. In a manner similar to
the $N=1$ case, the transition rate from each branch is estimated by
$\sim \frac{Ne^{-\beta V}}{3N-1}\frac{Ne^{-\beta V}}{1+Ne^{-\beta V}} = \frac{N^2e^{-\beta 2V}}{(3N-1)(1+Ne^{-\beta V})}$. Thus, the relaxation time of the
fluctuation of $p_1$ is estimated as the decrease with $N$ as
\begin{equation}
\tau^{L}_{d}(3) \sim  \frac{(3N-1)(1+Ne^{-\beta V})}{N^2}e^{2\beta V}.
\end{equation}
Considering the $e^{\beta V}$ dependence of $\tau_{N\rightarrow
\infty}$, the above estimate is consistent with Fig. 6(b).

For $S$ larger than $3$, the transition diagram becomes rather
complicated. However, a similar analysis should be possible to
estimate the prolongation in the relaxation time.

\section{V. Summary and discussions}

In the present paper, the slowing down of the relaxation in
reversible catalytic reaction networks induced by the smallness of
molecule number is investigated as a general property of catalytic
reaction networks. This prolongation of relaxation is a result of
bottlenecks in reactions; these appear due to the deficiency of
the catalyst required for a reaction. The number of molecules can
be so small that the number of catalysts becomes zero. In this
case, a pair of a substrate and the corresponding catalyst
molecule species can hardly exist simultaneously.  Such a
constraint makes it difficult to realize a specific configuration
necessary for the relaxation. The probability for realization is
given by $\exp(-\beta E_{bottle})$, with $E_{bottle}$ as the
corresponding energy barrier to realize such rare conditions, or
the sum of such energy barriers. This bottleneck energy is
generally different from the energy gap in the continuum limit
that is obtained from the rate equation (ordinary differential
equation). Thus, the relaxation time at a small molecule number
deviates from the continuum case by the factor $\exp(\beta \delta
E)$ with an appropriate effective energy difference, $\delta E$.

By considering the models of simple catalytic reaction networks
consisting of resource chemicals of $S$ species and the
corresponding products, we have demonstrated this deviation of
relaxation time from both direct simulations and analysis by using
Master equation. From the numerical and analytic estimates,
$E_{bottleneck}=2V$ and $\delta E=V$ for $S=3$, where $V$ is the
energy gap between the resource and the product chemicals. For
$S>2$, in general, the prolongation of the relaxation time becomes
prominent when $h=N\exp(-\beta V)$ is less than unity, and its
amplification ratio from the continuum limit is represented as a
function of $S$ and $h$. Note that the cascade system in the $N=1$
case is equivalent to the "Asymmetrically Constrained Ising Chain"
(ACIC), Hierarchically constrained Ising model, or East model,
which are studied as simple abstract models for glassy
states\cite{aici_1, aici_2, aici_3}. Following the interpretation
therein, the increase in relaxation time at $h < 1$ as a result of
the decrease in $N$ or temperature may be regarded as a type of
glass transition. According to the recent studies on ACIC, the
correlation time of the motion of $p_1$ (not the relaxation time
of the total system) is estimated as $\tau_1 \sim (1+2e^{\beta
V})^k$ where the integer $k$ obeys $2^{(k-1)} < S \le 2^k$
\cite{aici_2, aici_3}. In cases with $S=2, 3, 4$, this fact is
consistent with our estimate of the relaxation time of the cascade
system with $N=1$. The estimation of $\delta E$ as a function of
$S$ and $h$ for general cases both for cascade and loop systems is
an important issue that should be studied in the future.

In addition to the slow down in relaxation, the equilibrium
distribution deviates in a network called a loop system, where all
the reactions are catalyzed by one of the products. The constraint
that the numbers of a certain pair of chemical species cannot
simultaneously be zero leads to the deviation of the average
distribution of molecule numbers from the continuum limit. Again,
this deviation becomes prominent when $h$ is less than unity.

Although we have adopted simple network motives to analyze the
relaxation, the prolongation of relaxation time is quite general
in catalytic reaction networks. Catalytic bottlenecks often appear
as the number of molecules is decreased in a large variety of
reaction networks in which catalysts are synthesized
within\cite{SAK1,SAK2}. The present study can provide a basis for the
general case with complex networks, as the motives here are
sufficiently small and can exist within such complex networks.

Biochemical reactions generally progress in the presence of
catalysts that are themselves synthesized as products of such
reactions. These reactions form a network of a variety of chemical
species. Here, the molecule number of each species is generally
not very large. Hence, the slow relaxation process and deviation
from equilibrium discussed in this study may underlie
intracellular reaction processes. Moreover, the present network
motives are so simple that they are suggested to exist in
biochemical networks. We also note that the resource and product
in our model can be interpreted as non-excited and excited states
of enzymatic molecules. Indeed, many molecules are known to
exhibit catalytic activity only when they are in an excited state,
which can help other chemicals to switch to an excited state. In
fact, such networks with mutual excitation are known in
signal-transduction networks\cite{sig1,sig2,sig3}, where the
present slow relaxation mechanism may be relevant to sustain the
excitability of a specific enzyme type over a long time span. It
is important to pursue the relevance of the present mechanism in
cell-biological problems by considering more realistic models in
the future.

We also note that not only the discreteness in the molecule number
but also the negative correlation between a substrate and the
corresponding catalyst within a reaction network or in a spatial
concentration pattern suppresses the relaxation
process\cite{AK2,AK3,SAK1,SAK2}. The present mechanism due to
discreteness may work synergetically with the earlier mechanism to
further suppress the relaxation to equilibrium. The construction
of reaction networks to achieve slower relaxation together with
the network analysis will be an important issue in the future.

The authors would like to Shinji Sano, for informing us of his finding
on the prolongation of relaxation in reaction networks due to the
discreteness in molecule numbers, which triggers the present study.
A. A. was supported in part by a Grant-in Aid for Young Scientists
(B) (Grant No. 19740260).

\end{document}